# Demographics of Protoplanetary Disks: A Simulated Population of Edge-on Systems

Isabel Angelo[1,2], Gaspard Duchene[2,3], Karl Stapelfeldt[4], Zoie Telkamp[2,5], François Ménard[3], Deborah Padgett[4], Gerrit Van der Plas[3], Marion Villenave[4], Christophe Pinte[3,6], Schuyler Wolff[7], William J. Fischer[8], and Marshall D. Perrin[8]

[1] Department of Physics & Astronomy, University of California Los Angeles, Los Angeles, CA, 90095, USA
[2] Department of Astronomy, University of California, Berkeley, CA, 94720, USA
[3] Univ. Grenoble Alpes, CNRS, IPAG, F-38000 Grenoble, France
[4] Jet Propulsion Laboratory, California Institute of Technology, 4800 Oak Grove Drive, Pasadena, CA, 91109, USA
[5] Department of Astronomy, University of Virginia, P.O. Box 400325, Charlottesville, VA, 22904, USA
[6] School of Physics and Astronomy, Monash University, Clayton, Vic 3800, Australia
[7] Department of Astronomy and Steward Observatory, University of Arizona, Tucson, AZ, 85721, USA
[8] Space Telescope Science Institute, Baltimore, MD, 21218, USA
Received 2020 October 24; revised 2023 February 1; accepted 2023 February 8; published 2023 March 14

## Abstract

The structure of protoplanetary disks plays an essential role in planet formation. A disk that is highly inclined, or "edge-on," is of particular interest since its geometry provides a unique opportunity to study the disk's vertical structure and radial extent. Candidate edge-on protoplanetary disks are typically identified via their unique spectral energy distributions (SEDs) and subsequently confirmed through high-resolution imaging. However, this selection process is likely biased toward the largest, most-massive disks, and the resulting sample may not accurately represent the underlying disk population. To investigate this, we generated a grid of protoplanetary disk models using radiative transfer simulations and determined which sets of disk parameters produce edge-on systems that could be recovered by the aforementioned detection techniques—i.e., identified by their SEDs and confirmed through follow-up imaging with the Hubble Space Telescope. In doing so, we adopt a quantitative working definition of "edge-on disks" (EODs) that is observation driven and agnostic about the disk inclination or other properties. Folding in empirical disk demographics, we predict an occurrence rate of 6.2% for EODs and quantify biases toward highly inclined, massive disks. We also find that EODs are underrepresented in samples of Spitzer-studied young stellar objects, particularly for disks with host masses of $M \lesssim 0.5\,M_\odot$. Overall, our analysis suggests that several dozen EODs remain undiscovered in nearby star-forming regions, and provides a universal selection process to identify EODs for consistent, population-level demographic studies.

*Unified Astronomy Thesaurus concepts:* Protoplanetary disks (1300); Circumstellar disks (235); Planetary system formation (1257); Planet formation (1241); Radiative transfer (1335); Radiative transfer simulations (1967); Astronomical simulations (1857); Direct imaging (387)

## 1. Introduction

The exoplanet population exhibits a remarkable diversity of system architectures (e.g., Winn & Fabrycky 2015), which suggests a correspondingly diverse sample of the protoplanetary disks from which these planets form. Understanding the demographics of these disks can thus shed light on the formation mechanisms that contribute to exoplanet diversity. Observations of protoplanetary disks hold the key to this understanding as they offer a window into the relatively small time frame in which planets begin to form around their host stars. More specifically, the characteristic distributions of dust in these disks hold essential information about their structure, composition, and evolution and provide a detailed picture of the environments in which planets begin to form (Andrews 2020; Benisty et al. 2022).

A large number of such protoplanetary systems are known, with dozens already imaged at high angular resolution in the submillimeter with the Atacama Large Millimeter/submillimeter Array (ALMA) and/or in scattered light with the Hubble Space Telescope (HST) and ground-based adaptive optics instruments (Watson et al. 2007; Andrews 2020 and references therein). However, precise disk properties are difficult to uncover empirically from scattered-light images for systems at low-to-moderate inclinations due to the glare from the bright central star. In particular, despite the notable exception of HL Tau (Pinte et al. 2016), the vertical structure of protoplanetary disks and the relevance of dust settling in planet formation are difficult to constrain due to projection effects and generally require detailed modeling of the disk composition and configuration.

Highly inclined disks that block the observer's direct line of sight to the star help us address some of these issues because we can directly resolve their vertical structure and dust distributions at high altitudes. These disks, which we refer to as edge-on disks (EODs) irrespective of their exact inclination with respect to our line of sight, thus enable unparalleled studies of dust settling (e.g., Burrows et al. 1996; Stapelfeldt et al. 1998; Padgett et al. 1999). Observationally, we define these disks as systems that are optically thick at optical and near-infrared wavelengths along the line of sight to the central star, with stellar photons scattering off the surface layers of the disk before reaching the observer (Whitney & Hartmann 1992).

The sample of known EODs has grown significantly in the past ~20 yr. While the first few systems were discovered serendipitously, the uniform mapping of nearby star-forming regions across the near- to far-infrared with Spitzer (e.g., Evans et al. 2009) has facilitated systematic searches for new systems based on their characteristic spectral energy distributions (SEDs). In short, EODs appear as remarkably underluminous in the optical and near-infrared regimes compared to typical young low-mass stars at the same distance due to the presence of their opaque circumstellar disk. They also display copious mid- and far-infrared emission, a regime where dust opacity is sufficiently reduced and the line of sight to the inner disk becomes mostly optically thin (Stapelfeldt et al. 1997; Wood et al. 2002). This unique combination of observable features has been used to identify candidate EODs based on their SEDs prior to obtaining confirmation high-resolution imaging. For instance, building on Spitzer observations, Stapelfeldt et al. (2014) selected candidate EODs to be imaged at high resolution with HST, leading to the discovery of a dozen new such systems with an ≈50% success rate and bringing the sample of known EODs to about three dozen.

Assuming a population of randomly oriented disks and a standard height-to-radius aspect ratio, Stapelfeldt (2004) predicts ≈15% of protoplanetary disks to be observed as EODs based on their inclination and viewing geometry alone. This would imply the existence of a few hundred EOD systems in nearby star-forming regions, an order of magnitude more than the currently known sample. While it is possible that a number of EODs are still undiscovered, another explanation for the paucity of confirmed EODs is that geometry (specifically inclination) alone is insufficient to determine whether a disk is truly an EOD with our definition. For instance, it is possible that only the highest mass disks have sufficient opacity toward the central star to appear in scattered light when observed in the optical. It is also likely that the most-compact EODs remain unresolved even with HST imaging due to its limited angular resolution. If the majority of disks are compact, as suggest by ALMA observations (Long et al. 2019), the expected number of EODs could be much smaller than anticipated.

In this paper, we aim to transcend simple geometrical arguments and analyse the true demographics of disks that are confirmed as EODs. By assessing the biases inherent to the SED-selection methods used to identify candidate EODs, we can infer implications about the broader disk population from the current EOD sample. We simulate a population of protoplanetary disks that encapsulates a broad range of physical properties, such as total mass, radial and vertical extents, and inclination, to reflect the population of potentially observable disks. We then determine which ones would be characterized as "candidate" and "confirmed" EODs, based on their SEDs and optical images, respectively, and map which combinations of parameters produce edge-on (i.e., optically thick, for observational purposes) disks and which, if any, would be missed by an SED-based search strategy. We then interpret our findings by comparing the predicted and observed occurrences of EODs, as well as by analysing the observed distribution of stellar host properties.

We stress that our focus is on disks that can be imaged and confirmed as EODs with noncoronagraphic scattered-light images, best done with HST in the optical. Ground-based, near-infrared adaptive optics observations can also image some of these disks (e.g., Koresko 1998; Stapelfeldt et al. 1998; Monin & Bouvier 2000), but the complexity and variability of the point-spread function (PSF) introduces ambiguous interpretation of marginally resolved disks (e.g., Huélamo et al. 2010). Furthermore, most EODs are too faint to serve as guide stars, dramatically limiting the number of candidates that can be followed up even with extreme adaptive optics systems (e.g., SPHERE, GPI, SCExAO). For this analysis, we selected a WFC3/F606W observing mode as representative of HST imaging; differences between this mode and other cameras and filters on HST should be minor. In particular, while we expect more compact disks to be detectable with higher-resolution instruments like JWST at its shortest observing wavelengths or future ground-based extremely large telescopes (ELTs), all surveys to date have been conducted with HST or lower-resolution instruments, which motivates our analysis setup.

We begin by describing our grid of protoplanetary disk models in Section 2. In Section 3, we outline the "edge-on" criteria we use to map the disk parameters in our grid to candidate and confirmed EODs based on existing detection methods. In Section 4, we describe the empirically derived disk population that we simulate to infer the overall occurrence rate of EODs. Finally, we present quantified detection biases, predicted global disk occurrence rates for EODs, as well as their interpretation, in Section 5.

## 2. Model Grid

### 2.1. Modeling Framework

We generated a grid of disk models using the radiative transfer modeling code MCFOST (Pinte et al. 2006, 2009). We varied a number of disk properties in order to span a physical parameter space that encapsulates the observed protoplanetary disk demographics. The grid was therefore designed to be a coarse, yet comprehensive, sample of empirical disk properties.

The disks in our models are passively heated by a central host star. For each model, we compute the system's SED and a high-resolution (0″.02, or ≈3 au per pixel at a distance of 140 pc) scattered-light image at a wavelength of 0.6 μm, representative of HST observations of EODs. The SEDs are evaluated at 19 discrete wavelengths ranging from 0.55 μm to 2.7 mm, matching the typical sampling of existing observations of young stellar objects (YSOs). No circumstellar envelope or foreground extinction are included in our models.

Our model assumes a tapered-edge surface density profile for the disk structure: $\Sigma(R) = \Sigma_c (R/R_c)^\gamma \, e^{-\left(\frac{R}{R_c}\right)^{2+\gamma}}$, where $R$ is the radial distance to the host star along the midplane, $R_c$ is the so-called critical radius where the surface density profile transitions from a power law to an exponential, and $\gamma$ is the surface density exponent that governs the radial distribution of dust grains in the disk. This prescription for the surface density profile has been widely used in analysing submillimeter observations of protoplanetary disks (e.g., Andrews 2015).

Along the vertical direction, we assume a Gaussian density profile for the gas component, which is appropriate for hydrostatic equilibrium in a vertically isothermal disk. This information is reflected in the disk scale height, which is parameterized with a power law. We use $H(R) = H_0 \left(\frac{R}{R_0}\right)^\beta$, where $R_0 = 100$ au is an arbitrary reference radius and $\beta$ controls how flared the disk is. We consider two cases for the vertical distribution of the dust component, specifically with and without dust settling. Settling, which results from the drag dust grains experience as they orbit at Keplerian speeds within the pressure-supported, sub-Keplerian gas disk, affects grains differently depending on their size. High-resolution ALMA observations of disk confirm that millimeter-sized grains generally settle efficiently toward the midplane (Pinte et al. 2016; Villenave et al. 2020, 2022; Doi & Kataoka 2021), while optical and near-infrared scattered-light images indicate that submicron grains are well mixed up to the disk surface (e.g., Flores et al. 2021; Wolff et al. 2021). Turbulence in the disk, however, can counterbalance the drag force and lift even relatively large grains up to high elevations, depending further on the local gas density and, therefore, the distance to the star. In our modeling, we explore the range of possible outcomes with (1) a "no-settling" case, in which the dust is assumed to be well mixed with the gas throughout the disk, and (2) three "settling" cases, using the analytical model of Fromang & Nelson (2009) and varying the turbulence parameter, $\alpha$. Lower values of $\alpha$ correspond to more settled disks.

The sizes of individual dust particles ($a$) are distributed according to a power law, $\frac{dN}{da} \propto a^{-q}$, integrated over the entire disk. In models with settling, the local grain size distribution is effectively shallower (steeper) in the disk midplane (surface layers). The grain composition is set to astronomical silicates (Draine & Lee 1984) and compact, spherical grains are assumed so that Mie theory is used in the radiative transfer. Finally, we note that the interplay between phenomena such as dust settling, radial migration, and grain growth, and the ubiquitous presence of gaps and rings (e.g., Andrews et al. 2018; Long et al. 2019), make our parametric approach somewhat simplified. Nonetheless, we believe that the grid of models we created is effective for our purposes of exploring a wide range of disk properties on larger, population-sized scales.

### 2.2. Fixed Parameters

For the purpose of this study, a number of parameters remained fixed for all models in our grid. These parameters were chosen on the basis of minimizing the computation time while maintaining as diverse a set of disk models as possible. In other words, our goal was to allow variations in the components of the model that were most crucial to our study of EODs.

For computational purposes, we fix the disk outer radius, beyond which no dust can exist in our models, at $R_{out} = 600$ au. This value is well beyond the largest disks we model (see below), so the surface density at these distances is vanishingly small in all models. $R_{out}$ should thus have a negligible effect on our model SEDs and images.

We also fix the parameters of the grain size distribution. Specifically, we fix the minimum and maximum grain sizes to $a_{min} = 0.01$ $\mu$m and $a_{max} = 1$ mm, respectively, and the exponent to $q = 3.5$, typical of interstellar models (see, for example, Mathis et al. 1977; Draine & Lee 1984; Weingartner & Draine 2001), and commonly used in protoplanetary disk modeling (e.g., Woitke et al. 2016). Lastly, we assume a constant (integrated) gas-to-dust ratio of 1:100 for all models.

### 2.3. Varied Parameters

Our model grid explores eight key disk parameters: host-star mass ($M_\star$), disk-to-star mass ratio ($\frac{M_{gas}}{M_\star}$), critical radius ($R_c$), inner radius ($R_{in}$), surface density exponent ($\gamma$), flaring exponent ($\beta$), scale height ($H_0$), and dust settling (via the turbulence parameter, $\alpha$). The values sampled for each parameter are summarized in Table 1 and justified below.

The mass of the central star is varied in our grid to span the empirical distribution of masses for T Tauri stars effectively

**Table 1**
Varied Model Parameters

| Parameter | Values | Pop. Synthesis Weights[a] |
|---|---|---|
| Host-star mass ($M_\star$) [$M_\odot$] | 0.15, 0.3, 0.6, 1.2 | empirical |
| Disk-to-star mass ratio ($\frac{M_{gas}}{M_\star}$) | $10^{-4}$, $10^{-3}$, $10^{-2}$, $10^{-1}$ | log-uniform |
| Critical radius ($R_c$) [au] | 10, 30, 100, 300 | empirical |
| Flaring exponent ($\beta$) | 1.05, 1.15, 1.25 | uniform |
| Scale height ($H_0$)[b] [au] | 5, 10, 15, 20, 25 | model-based |
| Inner radius ($R_{in}$) [au] | 0.1, 1, 10 | … |
| Surface density ($\gamma$) | $-0.5$, $-1$ | … |
| Dust settling ($\alpha$) | $10^{-5}$, $10^{-4}$, $10^{-3}$, no settling | |
| Inclination ($i$) [°] | 15 values in [45–90] range | uniform in $\cos(i)$ |

**Notes.**
[a] See Section 4.
[b] Evaluated at $R_0 = 100$ au.

(e.g., Kroupa 2001). For each host star we compute the corresponding effective temperature using the 2 Myr isochrone from the PHOENIX-BTSettl evolutionary models (Allard et al. 2012), which is in turn used to calculate the passive heating of the model disk. We opted to vary the disk-to-star mass ratio rather than the disk mass itself because of the known correlation between disk and stellar mass (see the discussion in Andrews 2020), which we used to define the range of mass ratio values that we consider in this study.

The ranges of $R_c$, $R_{in}$, and $\gamma$ are based on empirical distributions (see Andrews 2015; Ansdell et al. 2016; Long et al. 2019; etc.). In particular, we selected a range of inner radii that extends from close to the sublimation radius ($\lesssim 0.1$ au for the selected stellar properties) to 10 au, typical of the large inner cavity of transition disks (Espaillat et al. 2014).

Following Woitke et al. (2010), we explored values of $\beta$ ranging from strongly flared ($\beta = 1.25$) to almost bow tie ($\beta = 1.05$). We also explored a range of disk scale heights based on the results of modeling studies of protoplanetary disks (e.g., Andrews 2015). Finally, the adopted range for the turbulence parameter is based on current theoretical expectations as well as observations of individual disks (Andrews 2020 and references therein).

For each set of the parameters described above, we generated a disk model at 15 inclinations ($i$) between 45° and 90° spaced uniformly in $\cos(i)$ as appropriate for isotropically oriented disks. Because none of our models produce EODs at 45°, we did not compute models at lower inclinations and assume that they will not lead to an EOD configuration. Overall, the grid contains 23,040 parameter combinations, each at 15 inclinations, composing a total of 345,600 configurations with distinct SEDs and 0.6 $\mu$m scattered-light images. All images are then convolved by an HST/WFC3 F606W PSF produced with `tinytim` (Krist 1995) and resampled to a pixel scale of 0″04 to match the properties of typical EOD images.

### 3. Model Edge-on Criteria

We now proceed to explore the model grid to assess which combinations of physical parameters and inclination result in an EOD configuration. Specifically, we assess whether a model has (1) an SED that makes it a candidate EOD in photometric surveys, and (2) a scattered-light image that would be unambiguously confirmed as such an object in high-resolution HST observations. Figure 1 shows SEDs and 0.6 $\mu$m images for a sample of inclinations for a representative grid model, which illustrate the criteria for characterizing models as detectably edge-on in both observables. In the development of these criteria, we consistently tested them against known systems and adjusted tunable parameters to achieve satisfying performance. A summary of the EOD observations used to validate our tests can be found in the Appendix.

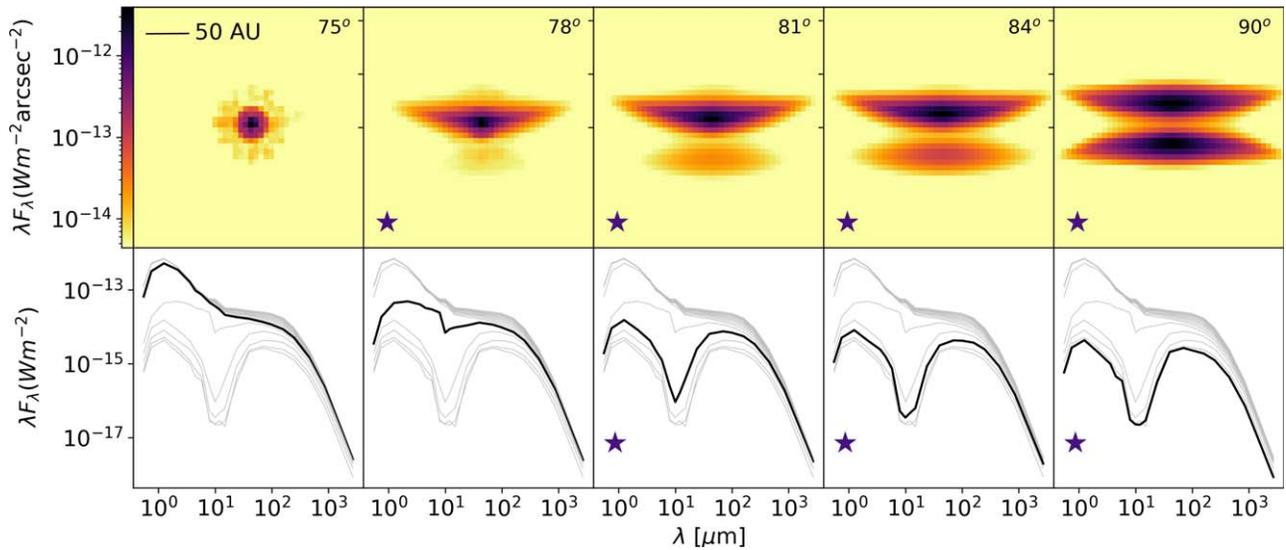

**Figure 1.** Example output for a grid model with the following parameters: $M_\star = 0.6\,M_\odot$, $\frac{M_{\rm gas}}{M_\star} = 10^{-1}$, $R_c = 100$ au, $\beta = 1.15$, $H_0 = 10$ au, $R_{\rm in} = 0.1$ au, $\gamma = -0.5$, and $\alpha = 10^{-3}$. The top and bottom panels show model 0.6 $\mu$m images and SEDs (both computed for a distance of 140 pc) at increasing inclinations from left to right. In the bottom row, the gray curves represent all 15 inclinations from 45° to 90°, while the black curve represent the specific inclination. SEDs and images that are considered edge-on by our criteria are denoted by a star symbol (see Section 3).

### 3.1. SED Criteria

We first consider which models have SEDs that would be classified as "candidate EODs" in photometric surveys. Since our models do not include foreground extinction, any model system that is significantly underluminous compared to the unobstructed central star is observed in scattered light. While this is an easy criterion to implement from a model perspective, it is not practical when considering observed populations, which contain a broad range of stellar luminosities and line-of-sight extinctions. We thus developed a robust set of criteria that a particular model must meet in order to be identified in our grid as a candidate EOD by its SED. To inform our selection criteria, We use a sample of known EOD systems for which we have SED and/or image data. The Appendix outlines the details of our sample and data processing methods.

For a model SED to be classified as edge-on, we require the SED to exhibit a combination of:

1. Significant starlight attenuation in the optical/near-infrared;
2. A near- to mid-infrared spectral slope typical of normal T Tauri stars; and
3. A strong mid- to far-infrared emission manifested as the disk transitions from optically thick to (mostly) optically thin to its own emission in the mid-infrared.

Figure 2 shows a schematic representation of these criteria, whose combination naturally gives rise to the characteristic double-peaked SED shape of EODs (e.g., Wood et al. 2002). In practice, we assess each criterion individually for each model +inclination combination, and assign a corresponding score of $0 \leqslant S \leqslant 1$, with higher scores indicating a higher likelihood of being classified as edge-on with existing SED-based search methods. As outlined below, we do not require that all three criteria are simultaneously satisfied but rather some combination thereof. This is to allow for the diversity of observed EODs, which cannot be grouped to form a single, tailor-made template.

The first test assesses whether the model SED is underluminous at 2.2 $\mu$m, where foreground extinction is negligible and pre–main-sequence stars that are not deeply embedded are visible. We evaluate this brightness deficit by computing the ratio between the observed and intrinsic stellar flux, $F_{2.2}/F_\star$. We then map this ratio to a corresponding edge-on score, $S_{2.2}$. Models with $F_{2.2}/F_\star \leqslant 0.1$ have $S_{\rm NIR} = 1$, corresponding to a high likelihood of being classified as edge-on, and models with $F_{2.2}/F_\star \geqslant 0.25$ are given an edge-on score of $S_{\rm NIR} = 0$ since they are too bright to be classified as edge-on. Models with $0.1 < F_{2.2}/F_\star < 0.25$ are given an intermediate score that increases linearly from 0 to 1 to avoid the edge effects of a pure binary test. The limits on the tests were selected to minimize false detection while ensuring strong ($\geqslant 75\%$) attenuation by the disk in the near-infrared regime.

For the second test, we assess whether the model has near- to mid-infrared colors similar to those of unobstructed T Tauri stars, as observed in the population of known EODs (see Figure 3). Specifically, we evaluate the spectral slope of the SED in the 2–8 $\mu$m range (which we denote as $\alpha_{2-8}$) through a least squares fit. We then map $\alpha_{2-8}$ to a second edge-on score, $S_{\rm color}$, where $\alpha_{2-8} \geqslant 0$ corresponds to $S_{\rm color} = 0$ and $\alpha_{2-8} \leqslant -0.5$ corresponds to $S_{\rm color} = 1$, with intermediate $\alpha_{2-8}$ values corresponding to $0 < S_{\rm color} < 1$. These limits were defined based on the range of observed colors for nonedge-on T Tauri stars (e.g., Luhman et al. 2010) and known EODs in our sample (Appendix).

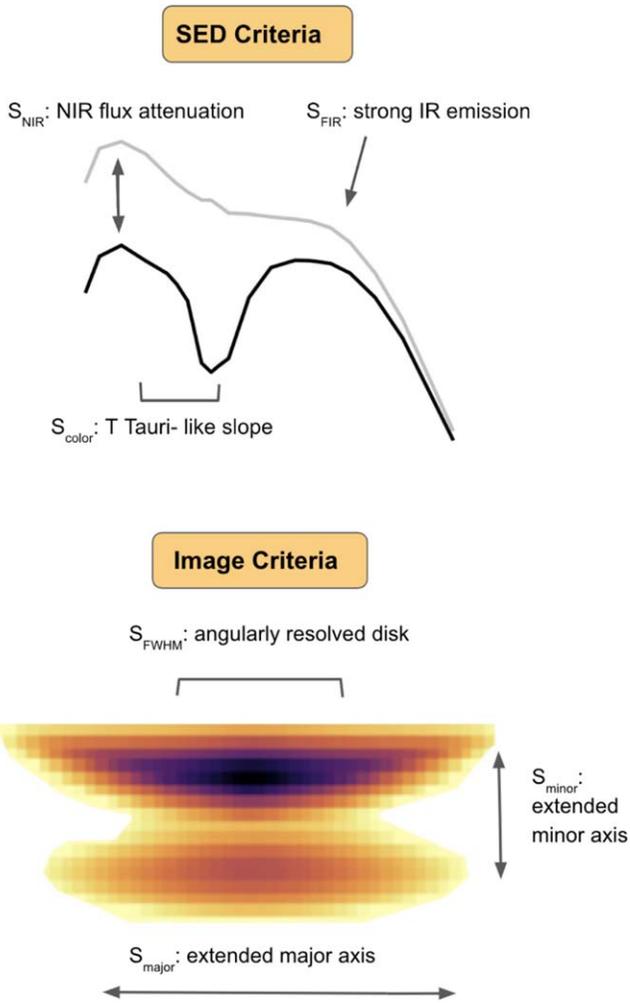

**Figure 2.** Schematic representation of our edge-on criteria for model SEDs (top) and images (bottom). For an SED, a combination of attenuated flux at near-infrared wavelengths ($S_{NIR}$), a T Tauri-like slope and color at mid-infrared wavelengths ($S_{color}$), and strong emission at far-infrared wavelengths ($S_{FIR}$) are needed to be classified as edge-on. For an image, the observed disk be angularly resolved with an FWHM larger than the image resolution ($S_{FWHM}$), and the disk flux should be extended along both the minor and major axes ($S_{minor}$, $S_{major}$).

This mapping effectively rejects embedded YSOs, or so-called Flat Spectrum and Class I sources (Lada 1987), which we collectively refer to as Class I sources in the remainder of this study. Due to their embedded nature, these objects are faint and bright at near- and far-infrared wavelengths, respectively, and often present a strong 10 μm silicate absorption feature, thus representing a common source of false positives for EOD selection. Among known EODs, only objects surrounded by significant leftover envelopes have $\alpha_{2-8} > 0$, for instance. It is possible, under certain circumstances, for some EODs around T Tauri stars to mimic such objects without actually hosting an envelope (e.g., Glauser et al. 2008), as our grid of models also confirms. Since such objects are generally not identified as candidate EODs (i.e., based exclusively on their SEDs), our second test satisfyingly excludes such models.

The third and final test determines whether the SED suggests the characteristic transition from optically thick to optically thin in the near- to far-infrared. This often (though not always) translates to a double-peaked shape around a mid-infrared trough for EOD systems (Stapelfeldt 2004), as can be seen in Figure 1. We quantitatively describe this behavior as high SED flux evaluated in the mid- to far-infrared relative to the near-infrared flux. To compute this, we evaluate the ratio of the maximum SED flux in the 15–100 μm range to the flux at 2.2 μm, $F_{FIR}/F_{NIR}$, and map this onto a third edge-on score, $S_{FIR}$. Models with $F_{FIR}/F_{NIR} \leqslant 0.25$ are considered to be too faint in the far-infrared to be classified as EODs, and are assigned $S_{FIR} = 0$. Models with $F_{FIR}/F_{NIR} \geqslant 0.5$ are optically thin to their far-infrared emission, consistent with EODs, and are given a score of $S_{FIR} = 1$. Similar to the other tests, models with intermediate values of $0.25 < F_{FIR}/F_{NIR} < 0.5$ are given an intermediate score of $0 < S_{FIR} < 1$. These limits were once again informed by the sample of known SEDs.

A summary of the three SED tests is presented in Table 2. Once we computed the associated scores for each of the three tests ($S_{2.2}$, $S_{color}$ and $S_{FIR}$), we assigned a comprehensive edge-on score to each model by averaging the three scores. We then classify models with an averaged overall score $S_{SED} \geqslant 0.7$ as candidate EODs, and the associated range of inclinations over which this condition is fulfilled is recorded. Our construction was designed to equally weight all three of our SED tests while requiring a nonzero score in all three categories to be classified as edge-on. This is to avoid incorrectly classifying low-inclination disks around very-low-mass stars or transition-disk systems as candidate EODs (both of which would pass two of the three tests but fail the third one).

While the tests presented above are designed based on known EODs and "typical" model EODs, we visually inspected the SEDs of a few tens of models across the entire grid at all inclinations and manually classified them as EOD or non-EOD. We found that the $S_{SED}$ score computed automatically was returning the same set of EODs as our visual inspection, confirming that it achieved the desired goal. When we applied the same tests to the population of known EODs (Appendix), we find that most objects that do not achieve $S_{SED} \geqslant 0.7$ are either embedded young stars, or systems in which the line of sight to the central star becomes optically thin at around 1–2 μm (e.g., IRAS 04158+2805, FS Tau B), which the tests presented here are not geared to encapsulate. Overall, we are confident that our methodology performs well to identify EOD candidates in the model grid, even though it could miss a few pathological cases.

### 3.2. Image Criteria

We developed a similar process to determine which of our models are "true EODs," i.e., systems that yield visibly edge-on images at 0.6 μm. Specifically, we developed three separate edge-on criteria (see Table 2) and combined them in a global test. Here we define a confirmed EOD as one whose image is:

1. Significantly angularly resolved;
2. Sufficiently extended along its major axis; and
3. Sufficiently extended along its minor axis.

A schematic representation of these criteria are shown in Figure 2. While images of most disks satisfy these conditions in principle, even at low inclinations, this is not always the case at low contrasts. Thus, we tailor these tests to be passed at the contrasts of HST (0.″04 per pixel) so that images where the bright central star is directly visible are not selected. As for the SED tests, the tests are designed to be effective on both synthetic and real images of EODs.

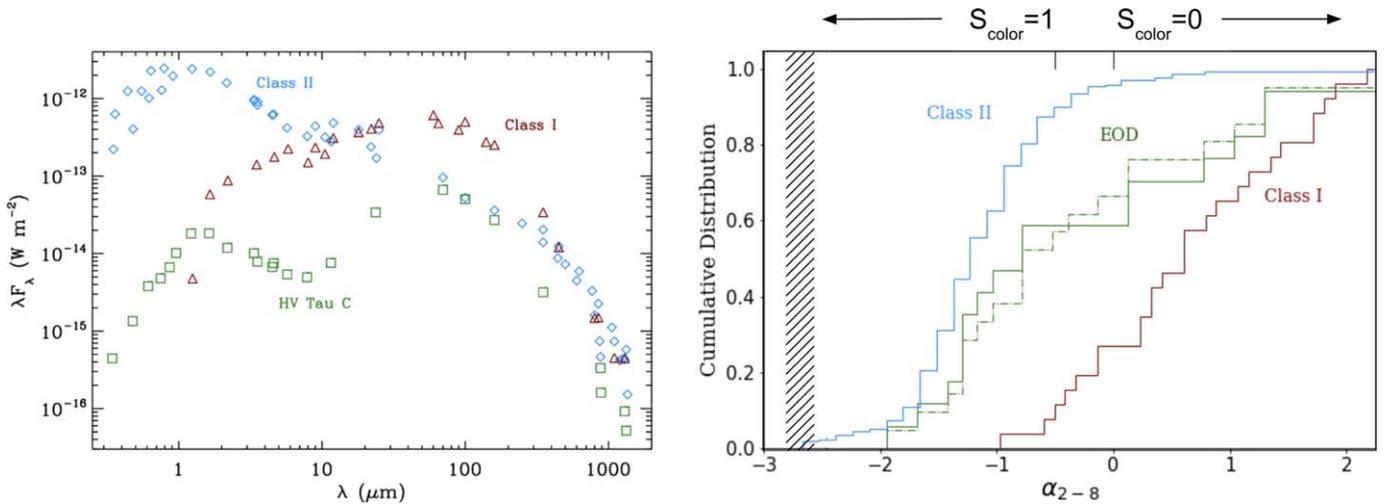

**Figure 3.** Left: SED of HV Tau C (Duchêne et al. 2010), a representative EOD with a well-sampled SED, compared to the Class I and II Taurus templates from Ellithorpe et al. (2019) and Ribas et al. (2017), respectively. Right: cumulative distribution of $\alpha_{2-8}$ for the Class I and Class II Taurus sources (Ellithorpe et al. 2019 and Luhman et al. 2010, respectively) compared to the observed distribution of EODs. The green dashed histogram represents the entire sample listed in Table 5 whereas the solid green histogram includes only objects whose SEDs are not affected by multiplicity, source confusion, or a significant envelope. For comparison, the spectral slope the photosphere model we used in our modeling lies within $\alpha_{2-8} = [-2.6, -2.8]$ for $M_\star = 0.15, 0.3, 0.6,$ and $1.2\,M_\odot$, as shown by the hatched region. The two tick marks at the top of the panel mark the lower and upper limit of the linear low-pass filter used to ascribe the $S_{\rm color}$ score.

The first test determines whether the object is angularly resolved. To this end, we fit a two-dimensional Gaussian to the model image and compute the ratio of the major axis FWHM to that of the PSF. We then use a high-pass filter, with $S_{\rm FWHM} = 0$ and 1 if this ratio is <1.5 and >2, respectively. Although the PSF stability of HST is exquisite and even slightly less-resolved objects could convincingly be distinguished from point sources, we require a relatively large value of the ratio to avoid confusion with compact envelopes or close binary systems.

The second test ensures that the disk is significantly extended along its major axis. Specifically, we compute the ratio of the brightest pixel in a column located ≈0″.3 (≈45 au at the assumed distance of 140 pc) from the center to the brightest pixel in the entire image. This ratio is fed into a high-pass filter to return scores, $S_{\rm major}$, of 0 and 1 if it is <0.02 and >0.4, respectively. These thresholds were designed to ensure that the disk surface brightness profile is substantially higher than that of the PSF at this small projected distance.

The third test is a similar surface brightness ratio but performed along the minor axis. When close to a 90° inclination, EODs present two similar brightness nebulae. In contrast, at lower inclinations the flux ratio between the nebulae increases while the vertical brightness profile gradually widens due to projection effects, as can be seen in Figure 1. To test for this in our models, we compute the ratio of the brightest pixel in a row located ≈0″.25 (≈35 au) above or below the center to the brightest pixel in the entire image. The high-pass filter for this ratio returns edge-on scores, $S_{\rm minor}$, of 0 and 1 if it is <0.05 and >0.1, respectively. As in the previous test, the thresholds are based on the brightness profile of the PSF.

As for the SED test, a final comprehensive score is computed as the average of all three scores described above, $S_{\rm image}$. The model is confirmed as being edge-on at a given inclination if $S_{\rm image} \geqslant 0.5$. This confirmation threshold was set to reflect the fact that not all models are angularly resolved in similar ways. As constructed, the test requires that the model be well resolved while still allowing for different image morphologies.

As is true for the SEDs, altering the final image test thresholds hardly affects our conclusions, mostly only changing the minimum inclination at which a model is deemed edge-on by one inclination bin. Visual inspection of a random sample of models confirmed that these thresholds are satisfactory. For the example model in Figure 1, these images become observably edge-on at $i \approx 78°$ with a probability of $S_{\rm image} = 0.67$ from $S_{\rm image} = 0$ at $i = 75°$. We also corroborate our image tests using images of known EODs from our sample. Unsurprisingly, we find that the vast majority of all known EODs have high-resolution images that pass the test we developed (see the Appendix for more information), with most exceptions being objects with bright, compact sources in the image center (e.g., IRAS 04200+2759).

### 4. Population Synthesis

Our model grid computes the *relative* EOD occurrence rate of simulated disks as a function of our varied grid parameters, which are sampled (log-)uniformly across the parameter space. In order to derive a *true* EOD occurrence rate for a population of young stars, we must also account for the fact that disks may not occur uniformly in all regions of our modeled parameter space. To do this, we simulate a population of $10^6$ disks informed by the underlying distribution of disk properties as reported in the literature. We then compute EOD occurrences of this simulated population using the criteria described in Section 3. Here we present how the grid parameter values are weighed to generate a synthetic population that we use to compute our nominal occurrence rates, though we explore variations on this model on Section 5.1.

#### 4.1. Inclination

The angular momentum vectors of disks are known to be isotropically distributed (e.g., Ménard & Duchêne 2004), for which our grid is designed in its sampling between 45° and 90°. To account for inclinations lower than 45° in this distribution, which are assumed to never be edge-on and are thus not included in our grid, we scale the overall probability

Table 2
Summary of the Edge-on Tests

| | Criterion | Measured Quantity | $S = 0$ Value[a] | $S = 1$ Value[a] |
|---|---|---|---|---|
| Edge-on SED | | | | |
| $S_{2.2}$ | Near-infrared starlight attenuation | $F_{2.2}/F_\star$ | 0.25 | 0.10 |
| $S_{\rm color}$ | T Tauri-like spectral slope | $\alpha_{2-8}$ | 0.0 | $-0.5$ |
| $S_{\rm FIR}$ | Strong far-infrared emission | $F_{\rm FIR}/F_{\rm NIR}$ | 0.25 | 0.50 |
| Edge-on Image | | | | |
| $S_{\rm FWHM}$ | Angularly resolved disk | $a_{\rm model}/a_{\rm PSF}$ | 1.5 | 2 |
| $S_{\rm major}$ | Extended disk along the major axis | $F_{\rm major}/F_{\rm sym}$ | 0.02 | 0.1 |
| $S_{\rm minor}$ | Extended disk along the minor axis | $F_{\rm minor}/F_{\rm sym}$ | 0.1 | 0.2 |

**Note.**
[a] These values define the range over which a model disk edge-on probability changes from 0 to 1. Models for which the measured quantity is outside this range are uniformly assigned scores of $S = 0$ or $S = 1$.

distributions down by 29.3%, which is the fraction of disks with $i = 0°–45°$ in an isotropic distribution.

### 4.2. Host Mass

For the underlying distribution of host-star masses in the population, we assume that the stars are distributed according to a Kroupa-type (Kroupa 2001) initial mass function (IMF). Specifically, we randomly sampled from that IMF[9] and binned the resulting stellar masses into four bins of log-equal width centered at the nominal values sampled in our grid. We found relative occurrences of 39%, 31%, 21%, and 9% from the lowest to the highest masses.

### 4.3. Disk-to-star Mass Ratio

To model the distribution of disk-to-star mass ratios, we build on the results of millimeter continuum surveys of nearby star-forming regions (e.g., Andrews et al. 2013; Barenfeld et al. 2016), which showed that the underlying distribution of the disk mass fraction in these regions is roughly log-uniform. We note that there may be a weak correlation between disk mass and host-star mass (as suggested by Pascucci et al. 2016). However, this correlation is still under debate, and a linear relationship between these parameters may be sufficient to describe existing observations (van der Marel & Mulders 2021). This motivated us to assume a log-uniform distribution in our analysis.

### 4.4. Flaring Exponent

From a theoretical standpoint, the assumption of hydrostatic equilibrium in disks yields flaring exponents between $\frac{9}{8}$ and $\frac{9}{7}$ under simple approximations (e.g., Kenyon & Hartmann 1987; Chiang & Goldreich 1997). The inclusion of settling in models reduces somewhat the flaring index (Chiang et al. 2001), although $\beta > 1$ remains a necessary requirement so that the outer upper surface of the disk is directly illuminated by the central star. This broad range is also consistent with observations (Burrows et al. 1996; Stapelfeldt et al. 1998; Avenhaus et al. 2018), although it must be emphasized that this quantity is difficult to disambiguate completely from other parameters. Given these elements, we decided to apply uniform weights over the values of $\beta$ used in our grid.

---

[9] https://github.com/keflavich/imf

### 4.5. Dust Settling

The dust settling in our model grid is dictated by the turbulent viscosity coefficient $\alpha$. Disk models generally consider a broad range of turbulence strengths, from strong to weak ($\alpha \gtrsim 10^{-2}$ to $\alpha \lesssim 10^{-4}$). Empirical observations, through SED fitting, spectral line width measurements, or analyses of resolved disk images, favor weaker turbulence levels (Mulders & Dominik 2012; Pinte et al. 2016; Flaherty et al. 2020; Trapman et al. 2020; Villenave et al. 2022). In most cases, this parameter remains poorly constrained, however, and we thus assume a log-uniform distribution for $\alpha$.

### 4.6. Critical Radius

The distribution of disk sizes has been constrained by recent high-resolution ALMA surveys (e.g., Long et al. 2019; Hendler et al. 2020). Although the completeness of existing surveys is uncertain, these provide a solid basis to use an empirically informed weighting scheme on this quantity. In addition to a large scatter from "compact" ($R_c \lesssim 40$ au) to very large ($R_c \gtrsim 200$ au) disks, a roughly linear correlation between disk size and stellar mass appears significant (Andrews 2020 and references therein). Most evident is the dearth of large disks among lower host-star masses ($M_\star \lesssim 0.3\,M_\odot$). Based on these facts, we adopt the following scheme in our analysis. We first assume a log-uniform distribution of $R_c$ for the highest stellar masses in our grid. We then randomly generate a distribution of disk sizes that is linearly scaled by each stellar mass and that we rebin to the same values as sampled in our grid. Our computed relative occurrence rates are plotted in the right panel of Figure 4. Interestingly, this procedure leads to fractions of disks with $R_c \lesssim 30$ au of 50%, 65%, 80%, and 95% for 1.2, 0.6, 0.3, and $0.15\,M_\odot$, respectively, which match well with the proportion of "compact disks" estimated by van der Marel & Mulders (2021). While this is not a perfect comparison, it confirms that the assumed distribution of $R_c$ is reasonably consistent with observations.

### 4.7. Scale Height

While the range of disk scale heights derived from studies of individual disks roughly matches that explored in our model grid (e.g., Burrows et al. 1996; Stapelfeldt et al. 1998), there are too few estimates available to define an empirical distribution. Instead we use our grid of models to compute a more physically informed distribution of $H_0$. Specifically, we use the midplane temperature computed as part of our radiative transfer calculations and convert it to a scale height assuming hydrostatic equilibrium (e.g., see their Sections 6.4 and 6.5 of Burrows et al. 1996). In order to treat all models equally, we perform this calculation at $R = R_c$ and then scale it according to the flaring exponent ($\beta$) to evaluate $H_0$ at $R_0$. This yields a new distribution of $H_0$ values, with each value corresponding to a single model in our grid. The resulting values are then binned to reflect our grid sampling of $H_0$.

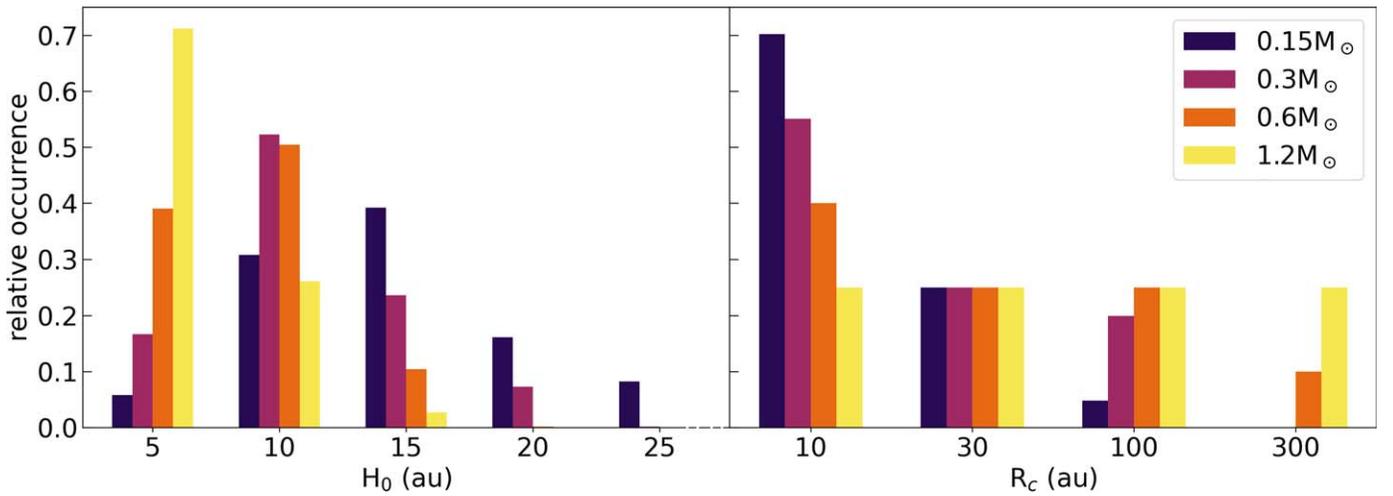

**Figure 4.** Left: distributions of $H_0$ (i.e., disk scale height evaluated at 100 au) for each stellar mass used in the population synthesis exercise, derived from the model-based midplane temperature. Right: similarly, distributions of $R_c$ for different stellar masses, based on the observed disk size distributions in ALMA observations and a linear ensemble correlation between stellar mass and disk size (see Section 4.6).

Crucially, this approach takes into account stellar mass and luminosity, the two key parameters that affect $H_0$ the most, albeit in opposite directions. All else being equal, a higher-mass star is more luminous, thus it heats up the disk more, but its gravity is also stronger. As Figure 4 illustrates, the latter effect is the most important within our grid, and the distribution of $H_0$ is skewed toward the lowest values for the higher-mass stars, consistent with the trends recovered by Walker et al. (2004) for brown dwarf circumstellar disks. We note that the inferred scale height may slightly underestimate the effective scale height as disks are not vertically isothermal. A higher temperature in a directly illuminated disk surface leads to a more-extended vertical profile than the Gaussian profile used here. Nonetheless, since disks are roughly isothermal up to ∼2 scale heights (e.g., Dullemond et al. 2007), we expect this to only have a small effect on our inferred disk occurrences.

## 5. Results and Implications

With the tests developed in Section 3, we proceed to determine the edge-on score for each model in the grid, corresponding to the likelihood that it would be classified as a "candidate" ($S_{\rm SED} \geqslant 0.7$) or "confirmed" ($S_{\rm SED} \geqslant 0.7$ and $S_{\rm image} \geqslant 0.5$) EOD. We then sample a large population of disks from the underlying population defined in Section 4, and map each disk in the simulated population to an EOD status (i.e., "candidate," "confirmed," or neither). This allows us to (1) evaluate the expected occurrence rate of EODs among entire populations of T Tauri stars, and (2) identify key trends in the physical parameter space that make certain disks over- or underrepresented among the existing population of known EODs.

### 5.1. EOD Occurrence Rates

From our simulated population of $10^6$ disks with underlying distributions outlined in Section 4, we compute an overall fraction of candidate EODs (i.e., using only the model SEDs) of 26.7%. Folding in the image-based edge-on probabilities, we find an occurrence rate of confirmed (i.e., using both model SEDs and images) EODs of 6.2%. Thus, only ≈1/4 of all candidate EODs would be confirmed by high-resolution imaging. This tendency is driven by the large frequency of low-mass and/or compact disks (see Garufi et al. 2018; Long et al. 2019) that are easily identified by their SEDs, but too small to be spatially resolved. Deeper, higher-resolution imaging, i.e., using the next generation of large ground-based telescopes, could confirm the true nature of some existing candidate systems. Likely or suspected examples of this category include Par-Lup 3-4 (Huélamo et al. 2010) and FW Tau C (Bowler et al. 2015), both of which are marginally resolved and appear to be significantly underluminous given their spectral type and/or dynamical mass. These are very likely compact EODs that HST images do not quite resolve.

Conversely, we find only a vanishingly small proportion (0.2%) of disks whose images are confirmed as edge-on but that would not be identified as candidates based on their SEDs. Many of these are borderline cases in the sense that increasing the disk inclination by just one step in our grid sampling (≈3°) is sufficient for the SEDs to go from not passing the EOD criteria to passing. In these cases, the lower inclination model provides enough optical depth in the optical to produce an EOD-like image, but not enough at 2.2 $\mu$m where the flux attenuation is evaluated. Others are objects with nearly flat SEDs resembling those of Class I sources, with little to no double-peaked shape.

In addition to our nominal candidate and confirmed EOD fractions, we also investigated the range of possible occurrence rates by varying the underlying assumptions of the population synthesis outlined in Section 4. In particular, we explore the affects of alternate weighting schemes for $R_c$, $\frac{M_{\rm gas}}{M_\star}$, and $H_0$, where our underlying assumptions are most uncertain. To start, it is possible that our $R_c$ includes too many large disks due to observational biases. Conversely, if gas emission extends much further out than submillimeter emission as predicted by Ansdell

Table 3
EOD Occurrence Rates

| Pop. Synthesis Weights | Candidate % | Confirmed % |
| --- | --- | --- |
| Default[a] | 26.7 | 6.2 |
| $R_c \geqslant 30$ | 23.3 | 13.5 |
| $R_c \leqslant 100$ | 27.3 | 5.8 |
| $\frac{M_{gas}}{M_\star} \leqslant 10^{-2}$ | 22.9 | 3.7 |
| $\frac{M_{gas}}{M_\star} \times 3$ | 28.5 | 7.0 |
| $H_0$, $T_c$ evaluated Above midplane | 34.0 | 8.4 |

**Note.**
[a] See Section 4. For all other rows, we specify the deviation from the default weights.

et al. (2018), then $R_c = 10$ au disks should be quite rare. To account for this, we consider the separate cases in which either $R_c = 10$ au or $R_c = 300$ au disks are excluded from our simulated population. For $\frac{M_{gas}}{M_\star}$, the disk mass fractions may be systematically underestimated due to the optically thin assumption that is often made to evaluate disk masses. Thus, we compute an upper limit on the magnitude of this affect by scaling up $\frac{M_{gas}}{M_\star}$ for all disks in our grid by a factor of 3. On the other hand, to account for the possibility that high-mass disks, which are close to gravitational instability and are rare, we compute occurrence rates for a population with $\frac{M_{gas}}{M_\star} = 0.1$ excluded. Finally, we account for potential underestimation of the disk scale height (see Section 4 for a description) by evaluating the temperature at $R = R_c$ at $3H_0$ above the disk midplane in our $H_0$ calculation, where the disk temperature is more characteristic of the $\tau = 1$ surface. The different values we compute for the EOD candidate and occurrence fractions under this range of assumptions are summarized in Table 3. We find a range of possible values for our candidate and confirmed occurrence rates of 23%–28% and 4%–8%, respectively, with two notable exceptions. Our alternate $H_0$ calculation predicts a candidate EOD occurrence rate of 34.0%, and removing $R_c = 10$ au predicts a confirmed occurrence rate of 13.5%. Outside of these, we find that the aforementioned variations produce <5% variations in our predicted occurrence rates.

We also investigated the effect that stellar binarity has on our predicted occurrence rates. In the event that a host has a stellar companion, the protoplanetary disk is predicted to be truncated to ∼1/3 of the binary semimajor axis (Artymowicz & Lubow 1994; Jang-Condell 2015). Thus, stellar binaries should have the most significant impact on the $R_c$ of our simulated population.[10] For simplicity, we estimate the maximum amplitude of this effect by looking at a subset of our grid with $M_\star = 1.2 \, M_\odot$, where the $R_c$ distribution is most heavily weighted toward large values and thus disk truncation due a companion is most significant (see Figure 4). We computed a new set of $R_c$ weights by simulating a population of stars using a binary fraction of 80% and the log-normal semimajor axis distribution reported in Raghavan et al. (2010). We draw $R_c$ for single stars in this population from the nominal uniform distribution for $M_\star = 1.2 \, M_\odot$ from Section 4. For the binaries, we compute $R_c$ to be 1/3 of the binary semimajor axis and exclude disks with $R_c$ that fall below our lowest grid bin, centered at $R_c = 10$. This procedure leads to 22%, 22%, 20%, and 35% of disks with $R_c = 10$, 30, 100, and 300 au, respectively. This distribution of $R_c$ values decreases the candidate occurrence rates of disks around $1.2 \, M_\odot$ host stars from 18.8% to 18.4%, and increases the confirmed occurrence rates from 10.3% to 10.7% for this subset, indicating that disk truncation by external companions is not a significant factor in predicting the occurrence rate of EODs.

### 5.2. Population-level Trends

Beyond the overall occurrence rate, our analysis reveals which types of disks are more or less likely to be candidate or confirmed EODs based solely on their physical parameters. Figure 5 illustrates the distribution of edge-on models marginalized over each variable parameter in our grid, as well as all two-dimensional distributions for each pair of parameters to reveal correlations between them. Several of the trends visible in this figure are unsurprising, but it is nonetheless interesting to consider them quantitatively. For example, while we expect a strong preference for disks with high inclinations among EODs, we find that models with an inclination of $i = 75°$ are only 3.4 times less likely to be edge-on than those inclined at 90°. This is contrary to the intuition that an inclination close to 90° is absolutely required for a disk to be edge-on with our definition. Indeed, several known EODs have inclinations that are lower than 80° (e.g., Krist et al. 1998; Glauser et al. 2008; Wolff et al. 2017).

Our findings also confirm predictions that EODs heavily favor models with a higher disk-to-star mass ratio since the optical depth through the disk is directly proportional to its mass. Our analysis quantifies this prediction: only about 7% (<1%) of models with disk-to-star mass ratios of $10^{-3}$ ($10^{-4}$) are confirmed EODs. Therefore, despite the fact that the two lowest disk-to-star mass ratios in our grid represent about half of all the disks in the synthesized global population, we expect a sample of bona fide EODs to be strongly biased toward high disk-to-star mass ratios.

Among the other parameters explored in this study, we find that the probability of a model being edge-on only has a shallow dependence, or no dependence at all, on the host mass, surface density index, and dust settling (see Figure 5, columns 2, 8, and 9, respectively) when all other parameters are fixed. A steeper dependence on host mass could have been intuitively expected given the strong correlation between disk and stellar mass. However, this is largely offset by the strong preference for compact disks and larger scale heights (Figure 4) among lower-mass stars, which effectively increases both their optical depth and geometrical thickness. Additionally, the relatively shallow effect of settling is worth noting. While we expected strongly settled disks to be less easily identified as edge-on given their reduced vertical extent, the small dust grains, which are responsible for absorption and scattering of visible stellar photons, are well mixed with the gas throughout the disk. This perfect coupling increases the EOD occurrence of disks with low-mass host stars, and only starts to break down at the smallest value of $\alpha$ explored in our grid.

The remaining parameters present more illuminating trends, particularly in the SED tests that determine whether a disk will be picked up as a candidate EOD. For instance, we find that the more compact the model disk is (i.e., the smaller $R_c$ is), the more likely it is to be identified as a candidate (see Figure 5, column 4). We stress that this trend shows how the candidate fraction varies with disk mass while holding all other parameters constant. For a fixed $R_c$, a higher disk mass causes higher volume densities and a higher total optical depth in the disk inner regions. However, if we consider only the confirmed EODs, i.e., taking into account HST imaging, models with $R_c = 10$ au are almost entirely absent (probability of being confirmed as edge-on of about 1%) since they are barely angularly resolved with HST, and even those with $R_c = 30$ au are significantly disfavored. Unsurprisingly, known EODs are biased toward the largest disks with $R_c \gtrsim 100$ au.

---

[10] Note that this effect applies only to binaries in which the disk is circumstellar. Analysis of the effect of binaries with circumbinary disks is beyond the scope of this paper.

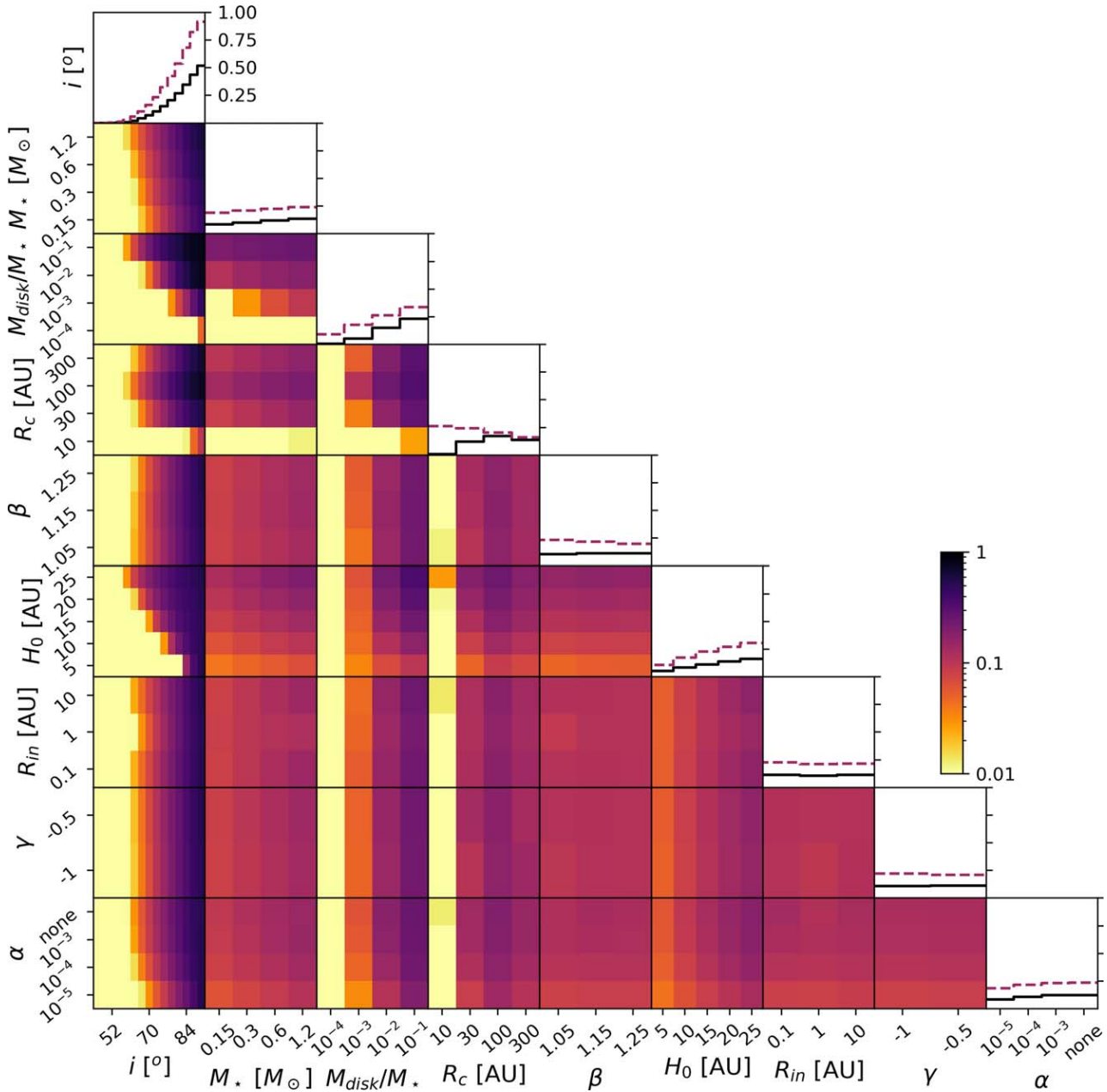

**Figure 5.** Corner plot illustrating the probability of being an EOD as a function of the model grid parameters. Along the diagonal, the probability distributions for "candidate" EODs ($S_{SED} \geqslant 0.7$) and "confirmed" ($S_{image} \geqslant 0.5$) are shown in dashed red and solid black, respectively. The nondiagonal panels show the "confirmed" edge-on fraction for a given combination of two parameters, where the color scale goes from 0 (pale yellow) to 1 (black).

We observe a similar behavior with the flaring index. We see from the top panel of Figure 5, column 5 that while disks are equally likely to be confirmed EODs for all values of $\beta$, less-flared disks (smaller $\beta$) are more likely to be candidate EODs. This may be counterintuitive, since more-flared disks are thicker in their outer regions and thus block light from the host star more efficiently, everything else being held constant. However, the inner region, where most of the opacity is concentrated, subtends a smaller angle for more-flared disks, reducing the inclination range over which a disk appears as a candidate EOD. This trend would be more prominent for candidate EODs than confirmed EODs because a larger degree of flaring leads to a more-extended scattered-light image, making the disk easier to detect and counterbalancing the aforementioned trend seen in the candidate EODs. We therefore deduce that samples of known EODs are not preferentially biased toward, or against, strongly flared or bow-tie disks.

Finally, we detect no significant trends associated with the disk inner radius, $R_{in}$. The edge-on fractions for the models fluctuate by only 0.5% for both $S_{SED}$ and $S_{image}$ as $R_{in}$ varies from 0.1 to 10 au (see Figure 5, column 7). This absence of dependence on $R_{in}$ implies that the EOD identification and confirmation processes are unbiased for (or against) transition disks, i.e., disks with large inner gaps. Given that transition disks are rather common in the general population (Ercolano et al. 2009; Espaillat et al. 2014), it is likely that a few of the known EODs are associated with transition-disk systems even though their nature cannot be confirmed by the usual methods. For instance, the large EOD associated with the very-low-mass IRAS 04158+2805 system is circumbinary in nature, with a ~200 au carved-out cavity (Ragusa et al. 2021). Indeed, this is the clearest example that neither the SED nor scattered-light images are sensitive to the disk inner radius in the edge-on configuration. It is worth noting in this context that IRAS 04158+2805 is one of the EODs whose inclination is farthest from 90°, thus providing a favorable viewing geometry. Had the disk been oriented closer to exactly edge-on, it is possible that its circumbinary nature would have remained hidden. Another example is FS Tau B, which was only found to have a large radial cavity after the system was imaged with ALMA (Villenave et al. 2020). There may be other EOD systems with similar cavities, but neither their SEDs nor scattered-light images can help assess this.

### 5.3. Discussion

The fraction of confirmed EODs predicted from our analysis is ≈6%, which is significantly lower than the simple geometrical argument initially presented by Stapelfeldt (2004). This is likely due to the dependence of the disk parameters on stellar mass, specifically the preponderance of low-mass stars and their associated low-mass disks. Either way, even this lower occurrence rate suggests that the current census of EODs is incomplete. Indeed, we would expect to find ≳60 confirmed EODs in a complete survey of nearby star-forming regions (given roughly a thousand disk-bearing young stars within ≈150 pc). This is significantly more than the two dozen EODs currently known (see Table 5). It is therefore very likely that there are several tens of undetected EODs in nearby star-forming regions, whose discovery and characterization would fill in missing gaps in the current demographic census.

To place our analysis in the context of the larger disk population better, we analysed the SEDs of known disk-bearing objects in the Taurus (Rebull et al. 2010), Chamaeleon, Lupus, and Ophiuchus (Dunham et al. 2015) clouds, which have been thoroughly studied with Spitzer and where the vast majority of known EODs reside. In each of these clouds, we selected all objects that had at least one solid (>5σ) detection in the MIPS bands and that were not classified as diskless Class III. We used the median K brightness to compute the $F_{2.2}/F_\star$ of all Class III sources in each region as representative of the typical stellar photosphere.[11] We then computed the $S_{SED}$ of these objects to determine the number of candidate EODs in these regions using our criteria from Section 3.1. As shown in Table 4, we find a consistent fraction of candidate EODs of about 10%,

---

[11] We acknowledge that this could allow some disk-bearing very-low-mass stars to appear artificially underluminous. However, with inconsistent spectral information for each target and potentially large line-of-sight extinctions, we lack a definitive method to assess the underluminous nature of any individual object and adopt this ensemble approach instead.

**Table 4**
Candidate EODs in Nearby Star-forming Regions

| Cloud | $N_{Objects}$ | $N_{Candidates}$ | $f_{Candidates}$ | References |
|---|---|---|---|---|
| Chamaeleon | 102 | 11 | $10.7^{+3.8}_{-2.3}\%$ | 1 |
| Lupus | 82 | 12 | $14.6^{+4.7}_{-3.1}\%$ | 1 |
| Ophiuchus | 263 | 24 | $9.1^{+2.1}_{-1.5}\%$ | 1 |
| Taurus (known)[a] | 140 | 11 | $7.9^{+2.9}_{-1.7}\%$ | 2 |
| Taurus (new)[b] | 40 | 7 | $17.5^{+7.5}_{-4.4}\%$ | 2 |
| Total | 627 | 65 | $10.4^{+1.3}_{-1.1}\%$ | |

**Notes.**
[a] Cloud members known prior to Rebull et al. (2010).
[b] Cloud members first reported in Rebull et al. (2010).
**References.** (1) Dunham et al. (2015) and (2) Rebull et al. (2010).

significantly lower than the 26.7% candidate fraction predicted by our population synthesis.

A likely explanation for this discrepancy is that the Spitzer catalogs of YSOs from which these disk-bearing objects are identified are incomplete. When Spitzer maps an entire star-forming region, the detected sources are filtered by magnitude and color selection criteria to avoid source confusion and isolate YSOs from more distant objects such as AGB stars and background galaxies (see, e.g., Rebull et al. 2010). As a result, at least some EODs, which are intrinsically underluminous, and other faint cloud members are likely excluded from the Spitzer YSO catalogs.

To investigate this, we considered the [4.5] ⩽ 11 selection criterion used by Rebull et al. (2010) as representative of the wavelength range over which the brightness filter was implemented for the Spitzer YSO catalogs. We evaluated which models in our grid would be excluded by these criteria and find that approximately half of our model EODs are too faint to be included in the Spitzer catalogs. This trend also exhibits a strong dependency on stellar mass, with 90% of the models on the low-mass end being cut and only 27% on the high-mass end. Thus, although the majority of high-host-mass EODs in our model grid may be well represented in the Spitzer surveys, many low-mass EODs are most likely filtered out. As a result of this bias, we would expect the sample of currently known EODs, which is in part built on the legacy of Spitzer surveys, to underestimate the true EOD fraction in the low-mass star regime severely. Figure 6 illustrates this using the Taurus population from Luhman (2018) as a reference. Strikingly, while half of the stellar population has a spectral type of M4 or later, only two such EODs are known, and only one is in Taurus. Thus, it appears that EODs are severely underrepresented in the current sample, largely as a consequence of the photometric selection criteria used to reduce confusion with unrelated background sources.

There are other factors that may also contribute to the discrepancy between our predicted EOD fraction and that suggested by Spitzer surveys. For example, it is possible that the Spitzer survey data are affected by contamination, which may not be fully accounted for by current selection criteria. In addition, due to the dependence of our SED criteria on some specific bandpasses, a single inaccurate or biased photometric measurement could be enough to lower the resulting estimate of $S_{SED}$ to below our detection threshold. Alternatively, it may be that our methodology systematically overpredicts the occurrence rate of EOD candidates. Indeed, we find that alternative distributions for $R_c$ and $\frac{M_{gas}}{M_\star}$ yield slightly lower candidate EOD fractions (see Section 5.1). Furthermore, since we find EOD occurrence fluctuations of nearly a factor of 2 for certain disk parameters (see Section 5.1), a small misrepresentation in the underlying disk population could affect our predictions. In particular, our treatment of the correlation of some parameters with stellar mass may be imperfect, or there may be additional unaccounted for correlations, e.g., between disk radius and disk mass.

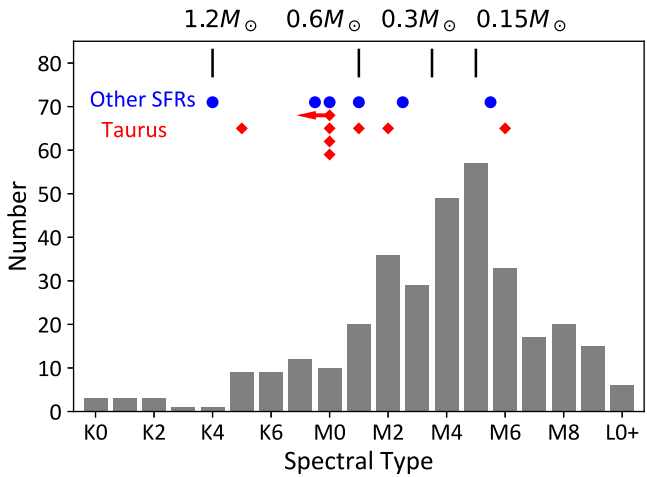

**Figure 6.** Distribution of spectral types for the entire Taurus population (gray histogram, from Luhman 2018) and for known nonembedded EODs (red diamonds and blue circles represent Taurus and non-Taurus EODs, respectively; see the Appendix). Black tick marks indicate the approximate spectral types for the four stars used in our model grid.

In any case, the fraction of our model EODs that fail the Spitzer YSO selection criteria, and its steep dependence on stellar mass, suggests that the occurrence rates of known EODs in nearby star-forming regions is incomplete, especially among low-mass stars. It is worth pointing out that these objects would likely have been detected by Spitzer observations and simply flagged out during the subsequent YSO target selection and it should be possible to design a new EOD-tailored criteria to retrieve them from global catalogs. Additionally, wide-field near-IR imaging with the future Roman Space Telescope, if selected, would be able to identify these missing EODs easily by their morphology. Thus, future studies to search for these EODs will be crucial for obtaining a more holistic sample of the underlying EOD population.

## 6. Conclusions

In this work, we generated SEDs and optical scattered-light images for a grid of protoplanetary disk models to investigate the demographics and any detection biases in the existing sample of EODs. We developed a working definition of "edge-on" that moves beyond a single inclination requirement to include disks that are optically thick and seen in scattered light at optical and infrared wavelengths. We used this definition to develop a set of criteria to classify disks as candidate and confirmed EODs based on their SEDs or images, respectively. We then determined from the model grid the combinations of physical parameters that produce detectable EODs. Finally, we simulated a population of disks based on their predicted underlying occurrences to determine the true occurrence rates of EODs as a function of disk and stellar properties.

We estimate global occurrence rates of 26.7% and 6.2% for candidate (based on their SEDs) and confirmed (based on their HST scattered-light images) EODs, respectively. The latter is roughly half of the expected occurrence rate that has been predicted from purely geometrical arguments. In addition to the expected biases toward higher inclinations and larger disk sizes, which we quantify for the first time here, we find that low disk-to-star mass ratios ($\leqslant 10^{-3}$), which represent up to half of all protoplanetary disks in star-forming regions, are highly unlikely to be observed as EODs. We further find that more compact disks are more likely to produce candidate EODs, although they often remain unresolved in scattered-light images. On the other hand, the surface density index ($\gamma$), dust settling ($\alpha$), and inner radius ($R_{in}$) have little to no influence on the probability of a disk to being detected as an EOD. The latter suggests that at least several known EODs are transition disks with large inner cavities.

Based on these occurrence rates, we conclude that the current census of EODs in nearby star-forming regions is significantly incomplete, likely due to survey sensitivities. In particular, we have determined that most EODs around very-low-mass stars would be too faint to pass the typical selection criteria used in Spitzer surveys to identify new YSOs in star-forming regions. If correct, up to several new dozen EODs could be identified with a dedicated high-resolution imaging survey with HST. Furthermore, higher-resolution instruments like JWST and upcoming ELTs will be sensitive to the most-compact missing disks, which we predict to be common. Beyond this, our analysis confirms that SED-based selection methods, initially designed based on a small sample of known EODs, are effective at identifying candidate EODs, although only about ≈1/4 of candidate EODs are confirmed to be true EODs from the images as a result of compact disks being difficult to resolve angularly.

A complete description of planet formation hinges on a robust understanding of EODs, which provide direct measurements of the vertical disk structure, temperature profile, and dust-settling environments from which planets are born. Additionally, a complete population of EOD observations will complement millimeter-wavelength disk observations to characterize further the dust settling properties in a larger number of disks. Our findings suggest that the current sample is both incomplete and overly biased toward high-mass disks, thus placing the existing sample of known EODs in the broader context of EOD occurrence and planet formation.

We would like to thank Tom Esposito, Paul Kalas, and the GPI team at UC Berkeley for insights on modeling, data processing, and analysis. I.A. and G.D. are partially supported by the NASA NNX15AD95G/NEXSS grant as well as the Graduate Deans Fellowship and Dorothy Radcliffe Dee Fellowship at UCLA. G.D. also acknowledges support from NASA grant 80NSSC18K0442. K.R.S. acknowledges support from HST GO grant 15148. M.V.'s research was supported by an appointment to the NASA Postdoctoral Program at the NASA Jet Propulsion Laboratory, administered by Oak Ridge Associated Universities under contract with NASA.

The Pan-STARRS1 Surveys (PS1) and the PS1 public science archive have been made possible through contributions by the Institute for Astronomy, the University of Hawaii, the Pan-STARRS Project Office, the Max Planck Society and its participating institutes, the Max Planck Institute for Astronomy, Heidelberg and the Max Planck Institute for Extraterrestrial Physics, Garching, The Johns Hopkins University, Durham University, the University of Edinburgh, the Queen's University Belfast, the Harvard-Smithsonian Center for Astrophysics, the Las Cumbres Observatory Global Telescope Network Incorporated, the National Central University of Taiwan, the Space Telescope Science Institute, the National Aeronautics and Space Administration under grant No. NNX08AR22G issued through the Planetary Science Division of the NASA Science Mission Directorate, the National Science Foundation grant No. AST-1238877, the University of Maryland, Eotvos Lorand University (ELTE), the Los Alamos National Laboratory, and the Gordon and Betty Moore Foundation.

The DENIS project has been partly funded by the SCIENCE and the HCM plans of the European Commission under grants CT920791 and CT940627. It is supported by INSU, MEN and CNRS in France, by the State of Baden-Württemberg in Germany, by DGICYT in Spain, by CNR in Italy, by FFwFBWF in Austria, by FAPESP in Brazil, by OTKA grants F-4239 and F-013990 in Hungary, and by the ESO C&EE grant A-04-046. Jean Claude Renault from IAP was the Project manager. Observations were carried out thanks to the contribution of numerous students and young scientists from all involved institutes, under the supervision of P. Fouqué survey astronomer resident in Chile.

This publication makes use of data products from the Two Micron All Sky Survey, which is a joint project of the University of Massachusetts and the Infrared Processing and Analysis Center/California Institute of Technology, funded by the National Aeronautics and Space Administration and the National Science Foundation," as well as the SIMBAD and Vizier databases, operated at CDS, Strasbourg, France

AllWISE makes use of data from WISE, which is a joint project of the University of California, Los Angeles, and the Jet Propulsion Laboratory/California Institute of Technology, and NEOWISE, which is a project of the Jet Propulsion Laboratory/California Institute of Technology. WISE and NEOWISE are funded by the National Aeronautics and Space Administration.

## Appendix
## EOD Sample

Table 5 lists all known EODs within ≈150 pc of the Sun considered in our analysis. We omit PDS 144 N (Perrin et al. 2006) from our sample as its central star is an intermediate-mass Herbig Ae star that is beyond the range of stellar properties considered in this study. We assembled the SED of each object from the Vizier database. This combines the Pan-STARRS, DENIS, UKIDSS, 2MASS, ALLWISE, AKARI, and IRAS all-sky surveys (Beichman et al. 1988; DENIS Consortium 2005; Skrutskie et al. 2006; Lawrence et al. 2007; Ishihara et al. 2010; Cutri et al. 2013; Doi et al. 2015; Chambers et al. 2016) with published and catalog photometry from pointed Spitzer, Herschel, and (sub-)millimeter observations. We systematically selected the highest-quality detections. IRAS 11030-7702 B is too embedded in the optical and near-infrared, while the SEDs of HK Tau B and LkHα 263 C are too contaminated by their brighter companions.

For the 19 systems with a well-sampled SED, we computed the edge-on score $S_{\rm SED}$ based on the tests described in Section 3. We can see from Table 5 that 63.6% of our selected known EODs pass our SED-based criterion. The nonnegligible fraction of systems that do not achieve a sufficient $S_{\rm SED}$ score is mostly a consequence of the fact that $S_{\rm NIR}$ is evaluated at 2 $\mu$m, whereas the disks are typically classified as EODs based on images taken in the optical. In some cases, the central star is hidden from direct view at the shorter wavelengths but peers through at near-infrared wavelengths. The other disks that are missed have a low $S_{\rm color}$, since their host star becomes directly visible within the 2–8 $\mu$m range, leading to a rising SED in the near- to mid-infrared range. In both cases, the SED of the system is markedly different from that of a "standard" EOD and tailoring tests to catch all of these systems is too system specific to be applied uniformly across many systems. We thus adopt the tests as defined in Section 3, which perform reasonably well overall.

We also applied the same series of tests to the actual images of known EODs. We excluded objects whose scattered light is clearly dominated by a circumstellar envelope rather than a disk (e.g., IRAS 04302+2247 or IRAS 04559+5200; Padgett & Stapelfeldt 1999; Sauter et al. 2009), since our models do not include such a component. In some cases, the best available observation of these EODs was not obtained with HST in the optical but from ground-based, near-infrared adaptive optics imaging. In all but one case, we employed the same method described in Section 3.2, although we scaled the angular distance from the center to compute $S_{\rm major}$ and $S_{\rm minor}$ in proportion to the FWHM of the PSF. In the case of IRAS 04158+2805, the largest disk in this sample (≈8″), the best image to analyse the entire structure of the disk is a seeing-limited image (Glauser et al. 2008), rather than existing HST images. Following the same recipe, we performed the major and minor axis brightness cuts at distances of 6.″7 and 5″, respectively, from the center. Of the 17 objects for which we could perform the tests, only two returned a value of $S_{\rm image}$ that is lower than 50%, i.e., it would not be classified as a confirmed EOD by our criteria. This is because these images are dominated by bright, very compact sources at the disk center. While visual inspection of these objects' HST images confirms the edge-on nature of these systems, the scattered-light nebulae are too faint to be picked up by the tests we constructed. Nonetheless, the overall success rate of the image-based criterion is 90.9%, a satisfyingly high value for us to proceed with our working definition of an EOD.

Table 5
Sample of Known EODs and Associated Edge-on Scores

| Object Name | SFR | Sp.T. | $S_{2.2}$ | $S_{color}$ | $S_{FIR}$ | $S_{SED}$ | $S_{FWHM}$ | $S_{major}$ | $S_{minor}$ | $S_{image}$ | Image Source[a] |
|---|---|---|---|---|---|---|---|---|---|---|---|
| Nonembedded Objects | | | | | | | | | | | |
| LkHα 263 C [1,2] | MBM 12 | M0[1] | ⋯ | ⋯ | ⋯ | ⋯ | 1 | 1 | 1 | 1 | ACS F555W [3] |
| FS Tau B [4] | Taurus | K5[5] | 1 | 0 | 1 | 0.67 | 1 | 1 | 1 | 1 | WFPC2 F555W [4] |
| HH 30 [6] | Taurus | M0[5] | 1 | 1 | 1 | 1 | 1 | 1 | 1 | 1 | WFPC2 F555W [6] |
| HK Tau B [7,8] | Taurus | M2[9] | ⋯ | ⋯ | ⋯ | ⋯ | 1 | 1 | 1 | 1 | WFPC2 F606W [7] |
| HV Tau C [10] | Taurus | M0[11] | 1 | 1 | 1 | 1 | 1 | 1 | 1 | 1 | WFPC2 F555W [12] |
| IRAS 04158+2805 [13] | Taurus | M6[5] | 0.37 | 0 | 1 | 0.46 | 1 | 1 | 1 | 1 | CFHTIR $I$ [13] |
| IRAS 04200+2759 [14] | Taurus | M2[15] | 0.97 | 0 | 1 | 0.66 | 1 | 0 | 0 | 0.33 | ACS F606W [16] |
| SSTtau J041941.4+271607 [16] | Taurus | ⋯[15] | 1 | 0.23 | 1 | 0.74 | 1 | 1 | 1 | 1 | ACS F606W [16] |
| SSTtau J042021.4+281349 [17] | Taurus | M1[17] | 1 | 0.11 | 1 | 0.70 | 1 | 1 | 1 | 1 | ACS F606W [16] |
| ESO-Hα 569 [18] | Chamaeleon | M2.5[19] | 1 | 1 | 1 | 1 | 1 | 1 | 1 | 1 | ACS F606W [16] |
| ESO-Hα 574 [14] | Chamaeleon | K8[19] | 1 | 1 | 1 | 1 | 1 | 1 | 1 | 1 | ACS F606W [16] |
| IRAS 11030-7702 B [14] | Chamaeleon | ⋯ | ⋯ | ⋯ | ⋯ | ⋯ | 1 | 1 | 1 | 1 | ACS F606W [16] |
| SSTc2d J160703.9-391111 [20] | Lupus | M5.5[21] | 1 | 1 | 1 | 0.67 | 1 | 1 | 0 | 0.66 | ACS F606W [22] |
| Flying Saucer [23] | Ophiuchus | M1[24][b] | 1 | 1 | 1 | 1 | 1 | 1 | 1 | 1 | WFPC2 F555W [3] |
| ISO-Oph 31 [25] | Ophiuchus | ⋯ | 0.96 | 0 | 1 | 0.65 | 1 | 1 | 1 | 1 | NACO $L'$ [25] |
| Oph E MM3 [26] | Ophiuchus | ⋯ | 1 | 0 | 1 | 0.67 | ⋯ | ⋯ | ⋯ | ⋯ | ⋯ |
| SSTc2d J162221.0-230402 [27] | Ophiuchus | ⋯ | 1 | 1 | 1 | 1 | 1 | 0 | 0 | 0.33 | ACS F606W [27] |
| SSTc2d J163131.2-242627 [14] | Ophiuchus | K4[28] | 1 | 1 | 1 | 1 | 1 | 1 | 0.11 | 0.70 | ACS F606W [16] |
| Embedded Objects | | | | | | | | | | | |
| DG Tau B [29] | Taurus | K5[30][b] | 1 | 0 | 1 | 0.67 | ⋯ | ⋯ | ⋯ | ⋯ | ⋯ |
| IRAS 04302+2247 [31] | Taurus | K0[32][b] | 1 | 1 | 1 | 1 | ⋯ | ⋯ | ⋯ | ⋯ | ⋯ |
| IRAS 04368+2557 [33] | Taurus | M2[34][b] | 1 | 0 | 1 | 0.67 | ⋯ | ⋯ | ⋯ | ⋯ | ⋯ |
| IRAS 04559+5200 [35] | CB 26 | M2[36][b] | 0.84 | 0.93 | 1 | 0.92 | ⋯ | ⋯ | ⋯ | ⋯ | ⋯ |

**Notes.**
[a] Discovery and image references: (1) Jayawardhana et al. (2002); (2) Chauvin et al. (2002); (3) HST Program 10603 (PI: D. Padgett); (4) Krist et al. (1998); (5) White & Hillenbrand (2004); (6) Burrows et al. (1996); (7) Stapelfeldt et al. (1998); (8) Koresko (1998); (9) Monin et al. (1998); (10) Monin & Bouvier (2000); (11) Appenzeller et al. (2005); (12) Stapelfeldt et al. (2003); (13) Glauser et al. (2008); (14) Villenave et al. (2020); Flores et al. (2021); Wolff et al. (2021); (15) Esplin & Luhman (2019); (16) Stapelfeldt et al. (2014); (17) Luhman et al. (2009); (18) Wolff et al. (2017); (19) Luhman (2007); (20) Ansdell et al. (2016); (21) Mortier et al. (2011); (22) HST Program 14212 (PI: K. Stapelfeldt); (23) Grosso et al. (2003); (24) Dutrey et al. (2017); (25) Duchêne et al. (2007); (26) Brandner et al. (2000); (27) Stapelfeldt et al. (2019); (28) Flores et al. (2021); (29) Stapelfeldt et al. (1997); (30) de Valon et al. (2020); (31) Lucas & Roche (1997); (32) Grafe et al. (2013); (33) Tobin et al. (2010); (34) Tobin et al. (2008); (35) Sauter et al. (2009); and (36) Launhardt & Sargent (2001). Observatories: HST (ACS, WFPC2, NICMOS), CFHT (CFHTIR), and VLT (NACO).
[b] Spectral type estimated from the system's dynamical mass.

## ORCID iDs

Isabel Angelo https://orcid.org/0000-0002-9751-2664
Gaspard Duchene https://orcid.org/0000-0002-5092-6464
Karl Stapelfeldt https://orcid.org/0000-0002-2805-7338
Zoie Telkamp https://orcid.org/0000-0001-6465-9590
François Ménard https://orcid.org/0000-0002-1637-7393
Deborah Padgett https://orcid.org/0000-0001-5334-5107
Gerrit Van der Plas https://orcid.org/0000-0001-5688-187X
Marion Villenave https://orcid.org/0000-0002-8962-448X
Christophe Pinte https://orcid.org/0000-0001-5907-5179
Schuyler Wolff https://orcid.org/0000-0002-9977-8255
William J. Fischer https://orcid.org/0000-0002-3747-2496
Marshall D. Perrin https://orcid.org/0000-0002-3191-8151

## References

Allard, F., Homeier, D., & Freytag, B. 2012, RSPTA, 370, 2765
Andrews, S. M. 2015, PASP, 127, 961
Andrews, S. M. 2020, ARA&A, 58, 483
Andrews, S. M., Huang, J., Pérez, L. M., et al. 2018, ApJL, 869, L41
Andrews, S. M., Rosenfeld, K. A., Kraus, A. L., & Wilner, D. J. 2013, ApJ, 771, 129
Ansdell, M., Williams, J. P., Marel, N. v. d., et al. 2016, ApJ, 828, 46
Ansdell, M., Williams, J. P., Trapman, L., et al. 2018, ApJ, 859, 21
Appenzeller, I., Bertout, C., & Stahl, O. 2005, A&A, 434, 1005
Artymowicz, P., & Lubow, S. H. 1994, ApJ, 421, 651
Avenhaus, H., Quanz, S. P., Garufi, A., et al. 2018, ApJ, 863, 44
Barenfeld, S. A., Carpenter, J. M., Ricci, L., & Isella, A. 2016, ApJ, 827, 142
Beichman, C. A., Neugebauer, G., Habbing, H. J., Clegg, P. E., & Chester, T. J. 1988, Infrared Astronomical Satellite (IRAS) Catalogs and Atlases. Volume 1: Explanatory Supplement, Vol. 1 (Pasadena, CA: NASA Jet Propulsion Laboratory)
Benisty, M., Dominik, C., Follette, K., et al. 2022, arXiv:2203.09991
Bowler, B. P., Andrews, S. M., Kraus, A. L., et al. 2015, ApJL, 805, L17
Brandner, W., Sheppard, S., Zinnecker, H., et al. 2000, A&A, 364, L13
Burrows, C. J., Stapelfeldt, K. R., Watson, A. M., et al. 1996, ApJ, 473, 437
Chambers, K. C., Magnier, E. A., Metcalfe, N., et al. 2016, arXiv:1612.05560
Chauvin, G., Ménard, F., Fusco, T., et al. 2002, A&A, 394, 949
Chiang, E. I., & Goldreich, P. 1997, ApJ, 490, 368
Chiang, E. I., Joung, M. K., Creech-Eakman, M. J., et al. 2001, ApJ, 547, 1077
Cutri, R. M., Wright, E. L., Conrow, T., et al. 2013, Explanatory Supplement to the AllWISE Data Release Products, 1, 1C
de Valon, A., Dougados, C., Cabrit, S., et al. 2020, A&A, 634, L12
DENIS Consortium 2005, yCat, II/263, 2263, 0D
Doi, K., & Kataoka, A. 2021, ApJ, 912, 164
Doi, Y., Takita, S., Ootsubo, T., et al. 2015, PASJ, 67, 50
Draine, B. T., & Lee, H. M. 1984, ApJ, 285, 89
Duchêne, G., Bontemps, S., Bouvier, J., et al. 2007, A&A, 476, 229
Duchêne, G., McCabe, C., Pinte, C., et al. 2010, ApJ, 712, 112
Dullemond, C. P., Hollenbach, D., Kamp, I., & D'Alessio, P. 2007, in Protostars and Planets V, ed. B. Reipurth, D. Jewitt, & K. Keil (Tucson, AZ: Univ. Arizona Press), 555

Dunham, M. M., Allen, L. E., Evans, N. J. I., et al. 2015, ApJS, 220, 11
Dutrey, A., Guilloteau, S., Piétu, V., et al. 2017, A&A, 607, A130
Ellithorpe, E. A., Duchene, G., & Stahler, S. W. 2019, ApJ, 885, 64
Ercolano, B., Clarke, C. J., & Robitaille, T. P. 2009, MNRAS, 394, L141
Espaillat, C., Muzerolle, J., Najita, J., et al. 2014, in Protostars and Planets VI, ed. H. Beuther et al. (Tuscon, AZ: Univ. Arizona Press), 497
Esplin, T. L., & Luhman, K. L. 2019, AJ, 158, 54
Evans, N. J., Dunham, M. M., Jørgensen, J. K., et al. 2009, ApJS, 181, 321
Flaherty, K., Hughes, A. M., Simon, J. B., et al. 2020, ApJ, 895, 109
Flores, C., Duchêne, G., Wolff, S., et al. 2021, AJ, 161, 239
Fromang, S., & Nelson, R. P. 2009, A&A, 496, 597
Garufi, A., Benisty, M., Pinilla, P., et al. 2018, A&A, 620, A94
Glauser, A. M., Ménard, F., Pinte, C., et al. 2008, A&A, 485, 531
Gräfe, C., Wolf, S., Guilloteau, S., et al. 2013, A&A, 553, A69
Grosso, N., Alves, J., Wood, K., et al. 2003, ApJ, 586, 296
Hendler, N., Pascucci, I., Pinilla, P., et al. 2020, ApJ, 895, 126
Huélamo, N., Bouy, H., Pinte, C., et al. 2010, A&A, 523, A42
Ishihara, D., Onaka, T., Kataza, H., et al. 2010, A&A, 514, A1
Jang-Condell, H. 2015, ApJ, 799, 147
Jayawardhana, R., Luhman, K. L., D'Alessio, P., & Stauffer, J. R. 2002, ApJL, 571, L51
Kenyon, S. J., & Hartmann, L. 1987, ApJ, 323, 714
Koresko, C. D. 1998, ApJL, 507, L145
Krist, J. 1995, in ASP Conf. Ser. 77, Astronomical Data Analysis Software and Systems IV, ed. R. A. Shaw, H. E. Payne, & J. J. E. Hayes (San Francisco, CA: ASP), 349
Krist, J. E., Stapelfeldt, K. R., Burrows, C. J., et al. 1998, ApJ, 501, 841
Kroupa, P. 2001, MNRAS, 322, 231
Lada, C. J. 1987, in Star Forming Regions, ed. M. Peimbert & J. Jugaku, Vol. 115 (Paris, France: IAU), 1
Launhardt, R., & Sargent, A. I. 2001, ApJL, 562, L173
Lawrence, A., Warren, S. J., Almaini, O., et al. 2007, MNRAS, 379, 1599
Long, F., Herczeg, G. J., Harsono, D., et al. 2019, ApJ, 882, 49
Lucas, P. W., & Roche, P. F. 1997, MNRAS, 286, 895
Luhman, K. L. 2007, ApJS, 173, 104
Luhman, K. L. 2018, AJ, 156, 271
Luhman, K. L., Allen, P. R., Espaillat, C., Hartmann, L., & Calvet, N. 2010, ApJS, 186, 111
Luhman, K. L., Mamajek, E. E., Allen, P. R., & Cruz, K. L. 2009, ApJ, 703, 399
Mathis, J. S., Rumpl, W., & Nordsieck, K. H. 1977, ApJ, 217, 425
Ménard, F., & Duchêne, G. 2004, A&A, 425, 973
Monin, J. L., & Bouvier, J. 2000, A&A, 356, L75
Monin, J. L., Menard, F., & Duchene, G. 1998, A&A, 339, 113
Mortier, A., Oliveira, I., & van Dishoeck, E. F. 2011, MNRAS, 418, 1194
Mulders, G. D., & Dominik, C. 2012, A&A, 539, A9
Padgett, D. L., Brandner, W., Stapelfeldt, K. R., et al. 1999, AJ, 117, 1490
Padgett, D. L., & Stapelfeldt, K. R. 1999, Comparison of OVRO Millimeter Array and/HST/NICMOS Images of the IRAS04302+2247 Circumstellar Disk (Washington, D.C.: Associated Universities, Inc.), 47
Pascucci, I., Testi, L., Herczeg, G. J., et al. 2016, ApJ, 831, 125
Perrin, M. D., Duchêne, G., Kalas, P., & Graham, J. R. 2006, ApJ, 645, 1272
Pinte, C., Dent, W. R. F., Ménard, F., et al. 2016, ApJ, 816, 25
Pinte, C., Ménard, F., Duchêne, G., et al. 2009, in AIP Conf. Ser. 1094, 15th Cambridge Workshop on Cool Stars, Stellar Systems, and the Sun, ed. E. Stempels (Melville, NY: AIP), 401
Pinte, C., Ménard, F., Duchêne, G., & Bastien, P. 2006, A&A, 459, 797
Raghavan, D., McAlister, H. A., Henry, T. J., et al. 2010, ApJS, 190, 1
Ragusa, E., Fasano, D., Toci, C., et al. 2021, MNRAS, 507, 1157
Rebull, L. M., Padgett, D. L., McCabe, C.-E., et al. 2010, ApJS, 186, 259
Ribas, Á., Espaillat, C. C., Macías, E., et al. 2017, ApJ, 849, 63
Sauter, J., Wolf, S., Launhardt, R., et al. 2009, A&A, 505, 1167
Skrutskie, M. F., Cutri, R. M., Stiening, R., et al. 2006, AJ, 131, 1163
Stapelfeldt, K. 2004, in IAU Symp. 202, Planetary Systems in the Universe, ed. A. Penny (Cambridge: Cambridge Univ. Press), 291
Stapelfeldt, K., Padgett, D., Burrows, C., Krist, J. & WFPC2 Science Team 1997, AAS Meeting Abstracts, 191, 05.15
Stapelfeldt, K., Padgett, D., Duchene, G., et al. 2019, AAS/Division for Extreme Solar Systems Abstracts, 51, 322.03
Stapelfeldt, K. R., Duchêne, G., Perrin, M., et al. 2014, in IAU Symp. 299, Exploring the Formation and Evolution of Planetary Systems, ed. M. Booth, B. C. Matthews, & J. R. Graham (Cambridge: Cambridge Univ. Press), 99
Stapelfeldt, K. R., Krist, J. E., Ménard, F., et al. 1998, ApJL, 502, L65
Stapelfeldt, K. R., Ménard, F., Watson, A. M., et al. 2003, ApJ, 589, 410
Tobin, J. J., Hartmann, L., Calvet, N., & D'Alessio, P. 2008, ApJ, 679, 1364
Tobin, J. J., Hartmann, L., & Loinard, L. 2010, ApJL, 722, L12
Trapman, L., Rosotti, G., Bosman, A. D., Hogerheijde, M. R., & van Dishoeck, E. F. 2020, A&A, 640, A5
van der Marel, N., & Mulders, G. D. 2021, AJ, 162, 28
Villenave, M., Menard, F., Dent, W. R. F., et al. 2020, A&A, 642, A164
Villenave, M., Stapelfeldt, K. R., Duchêne, G., et al. 2022, ApJ, 930, 11
Walker, C., Wood, K., Lada, C. J., et al. 2004, MNRAS, 351, 607
Watson, A. M., Stapelfeldt, K. R., Wood, K., & Ménard, F. 2007, in Protostars and Planets V, ed. B. Reipurth, D. Jewitt, & K. Keil (Tucson, AZ: Univ. Arizona Press), 523
Weingartner, J., & Draine, B. T. 2001, ApJ, 548, 296
White, R. J., & Hillenbrand, L. A. 2004, ApJ, 616, 998
Whitney, B. A., & Hartmann, L. 1992, ApJ, 395, 529
Winn, J. N., & Fabrycky, D. C. 2015, ARA&A, 53, 409
Woitke, P., Min, M., Pinte, C., et al. 2016, A&A, 586, A103
Woitke, P., Pinte, C., Tilling, I., et al. 2010, MNRAS, 405, L26
Wolff, S. G., Duchêne, G., Stapelfeldt, K. R., et al. 2021, AJ, 161, 238
Wolff, S. G., Perrin, M. D., Stapelfeldt, K., et al. 2017, ApJ, 851, 56
Wood, K., Lada, C. J., Bjorkman, J. E., et al. 2002, ApJ, 567, 1183